\documentclass[aps,preprint]{revtex4-1}
\usepackage{amsmath}
\usepackage{graphicx}
\usepackage{epstopdf}

\begin{document}

\title{ Rare Radiative $B_{c}\rightarrow D_{s1}(2460)\gamma$ Transition in QCD}
\author{K. Azizi$^{1}$ \footnote{e-mail: kazizi @ dogus.edu.tr}, N. Ghahramani$^{2}$  \footnote{e-mail: ghahramany @ susc.ac.ir}, A. R. Olamaei$^{2}$ \footnote{e-mail: olamaei@gmail.com}}
\affiliation{$^1$Physics Department, Faculty of Arts and Sciences, Do\u gu\c s University, Ac{\i}badem-Kad{\i}k\"{o}y, 34722 Istanbul, Turkey\\
$^2$Physics Department, College of Sciences, Shiraz University, Shiraz 71454, Iran\\
}
\begin{abstract}
We investigate the radiative $B_{c} \to D_{s1} \gamma$ transition in the framework of QCD sum rules.
In particular, 
 we calculate the transition form factors responsible for this decay in both weak annihilation and electromagnetic penguin channels using the quark condensate,
 mixed and two-gluon condensate diagrams as well as  propagation of the soft quark in the electromagnetic field as non-perturbative corrections.
These form factors  are then used to estimate the branching
ratios of the channels under consideration.
The total branching ratio of the $B_{c} \to D_{s1} \gamma$ transition is obtained to be in  order of $10^{-5}$, and the dominant contribution comes from the weak annihilation channel.

~~~PACS numbers: 11.55.Hx, 13.20.-v, 13.20.He
\end{abstract}

\maketitle

\section{Introduction}
The $B_c$ is the only heavy meson consisting of two heavy quarks with different flavors, hence the decay properties of this meson are of special interest. The
difference in heavy quark flavors forbids annihilation of this meson into gluons,  so the
excited $B_c$  states  undergo pionic
or radiative  transition to the pseudoscalar (PS) ground state when these states lie below  the threshold of the decay into the pair of heavy $B$ and $D$ mesons. The resulting PS ground state is  more stable compared to the corresponding quarkonia  and
decays mostly weakly. Because of this phenomenon, 
it is expected that the experimental study of
the $B_c$ meson and its decay properties will constitute an important part of the physics program at LHCb. The study of the heavy mesons will not only  provide  a window in
 extracting the most accurate values of the Cabbibo-
Kobayashi- Maskawa (CKM) matrix elements as the sources of the $CP$-violation in the Standard Model (SM)  but also will help us better understand  the perturbative and non-perturbative aspects of QCD.

In the present study, we work out the rare radiative $B_{c}\rightarrow D_{s1}(2460)\gamma$ transition in the framework of the QCD sum rules \cite{SVZ,Colangelo2}. Here, the $D_{s1}(2460)$ is the axial vector charmed-strange meson with
quantum numbers $J^P=1^{+}$ and the interpolating current $\eta_\nu=\overline{s}\gamma_{\nu}\gamma_{5}c$.  This transition proceeds via  both weak annihilation (WA) and
 electromagnetic penguin (EP)  of  flavor changing neutral current (FCNC) transition, based on the $b\rightarrow s\gamma$ at quark level. 
 We calculate the transition form factors responsible for this decay in both WA and EP modes using the quark condensate, mixed and two-gluon condensate diagrams, as well as
propagation of the soft quark in the electromagnetic field as non-perturbative corrections. We then use these form factors
 to estimate the  branching
ratios in both modes as well as the  total branching fraction of the $B_{c}\rightarrow D_{s1}(2460)\gamma$ transition. As  expected, the dominant contribution comes from the weak annihilation channel. 
Note that similar decays like the $B_{c}\rightarrow D^*_{s}\gamma$
 transition have been studied in the same framework \cite{azizi}.
 Some other radiative channels of the  $B_{c}$ meson like,
 $B_{c} \to l\overline{\nu}\gamma$
and $B_{c} \to B_{u}^{*}\gamma$ have also been previously studied using the QCD sum rules technique \cite{Aliev1,Aliev6}. For analysis of  other decay channels of the $B_{c}$ meson see, for instance,
\cite{Azizi2,Khosravi,kisibey1,kisibey2}.

The outline of the paper is  as follows.  In  Section II, we consider the radiation of the photon from both $B_{c}$  and $D_{s1}$ mesons, to construct the transition amplitude for the WA channel
in terms
of four relevant form factors. Two of the form factors $F_{V}^{(B_{c})}$ and $F_{A}^{(B_{c})}$, responsible for the emission of the photon from the initial state, are calculated in \cite{Aliev1}, and the remaining
two form factors $F_{V}^{(D_{s1})}$ and $F_{A}^{(D_{s1})}$, representing the emission of the photon from $D_{s1}$ meson, are
 calculated in Section III . In Section IV, we consider the two gluon condensate contributions, to calculate the transition form factors responsible for the EP mode.
 Finally, Section  V is devoted to the numerical analysis of the form factors and , calculation of the decay rates and branching ratios for the modes under consideration.
We also present results for  the total decay rate and
branching ratio of the $B_{c}\rightarrow D_{s1}(2460)\gamma$ transition. This Section also contains our concluding remarks.

\section{WEAK ANNIHILATION AMPLITUDE\label{Bound}}

In this section, we construct the  WA  amplitude    for the radiative  $B_{c} \to D_{s1} \gamma$ transition. Considering the quark contents of the initial and final mesonic states,
the possible diagrams are shown in figure 1.
\begin{figure}[tbp]
\centering {\includegraphics[width=15cm]{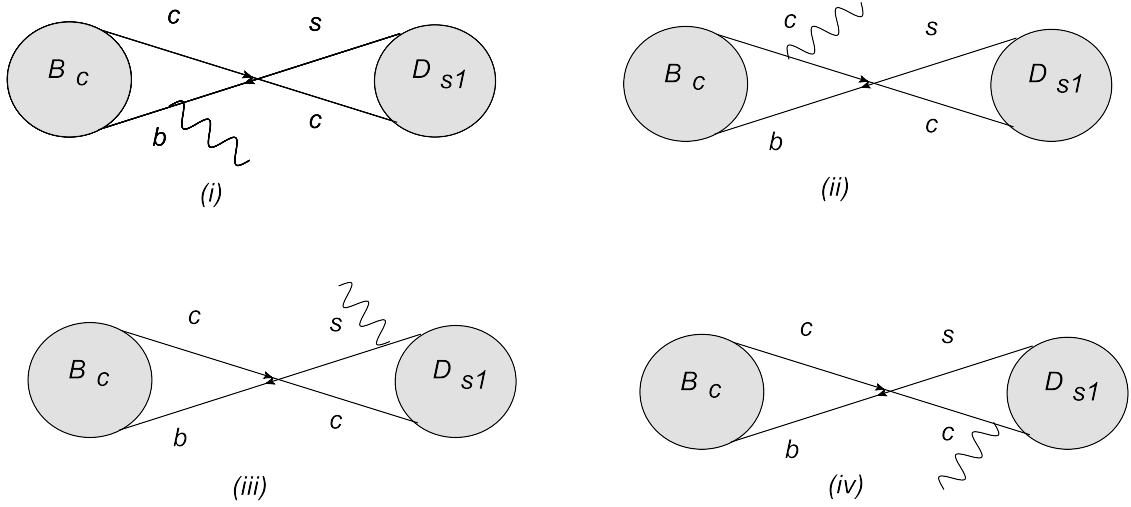}}
\caption{The weak annihilation mechanism for $B_{c} \to D_{s1} \gamma$.}
\label{Fig1}
\end{figure}
 Taking into account these diagrams, the transition amplitude for the radiative  decay under consideration  is  written as
\begin{equation}
M^{WA}(B_{c}\rightarrow D_{s1}\gamma)=\frac{G_{F}}{\sqrt{2}}V_{cb}V_{cs}^{*}\langle D_{s1}(p)\gamma(q)|(\overline{s}\Gamma_{\nu}c)(\overline{c}\Gamma^{\nu}b)|B_{c}(p+q)\rangle,  \label{weakamp}
\end{equation}
where $G_{F}$ is the Fermi weak coupling constant, $V_{ij}$ are elements of the CKM matrix,  $\Gamma_{\nu}=\gamma_{\nu}(1-\gamma_{5})$; and $p$, $q$ and $p+q$ are the momenta of the $D_{s1}$ meson,
photon and $B_{c}$ meson, respectively. To proceed further, we use the factorization hypothesis
and write the transition matrix element in Eq.(\ref{weakamp}) as
\begin{equation}
\langle D_{s1}(p)\gamma(q)|(\overline{s}\Gamma_{\nu}c)(\overline{c}\Gamma^{\nu}b)|B_{c}(p+q)\rangle=-e\varepsilon^{\mu}\varepsilon^{(D_{s1})\nu}
f_{D_{s1}}m_{D_{s1}}T^{(B_{c})}_{\mu\nu}-ie\varepsilon^{\mu}(p+q)^{\nu}f_{B_{c}}T^{(D_{s1})}_{\mu\nu}, \label{2}
\end{equation}
where we have divided the matrix element into two separate parts: the emission of the photon from the  $B_{c}$ meson (diagrams (i) and (ii) in figure 1), represented by the covariant tensor
$T_{\mu\nu}^{(B_{c})}$ and the emission of the photon from  the $D_{s1}$ meson denoted by the tensor $T_{\mu\nu}^{(D_{s1})}$ (see diagrams (iii) and (iv) in figure 1).
In Eq.(\ref{2}),  $f_{B_{c}}$ ($f_{D_{s1}}$) is the  decay constant of the $B_{c}$  ($D_{s1}$) meson and  $\varepsilon^{\mu}$ ($\varepsilon^{(D_{s1})\nu}$) is  the polarization
vector of the photon ($D_{s1}$ meson). The covariant tensors $T_{\mu\nu}^{(B_{c})}$ and  $T_{\mu\nu}^{(D_{s1})}$ are defined as
\begin{eqnarray}
T_{\mu\nu}^{(B_{c})}(p,q)\equiv i \int d^4xe^{iqx}\langle0|T\Bigg\{j_\mu^{em}(x)\overline{c}(0)\Gamma_{\nu}b(0)\Bigg\}|B_{c}(p+q)\rangle,
\end{eqnarray}
\begin{eqnarray}
T_{\mu\nu}^{(D_{s1})}(p,q)\equiv i \int d^4xe^{iqx}\langle D_{s1}(p)|T\Bigg\{j_\mu^{em}(x)\overline{s}(0)\Gamma_{\nu}c(0)\Bigg\}|0\rangle,
\end{eqnarray}
where $j_\mu^{em}$ is the electromagnetic current and $T$ is the time ordering operator. Applying the Ward identity for the electromagnetic current,  using  $q^2=0$ for the real photon, 
$\varepsilon.q=0$ and $\varepsilon^{(D_{s1})}.p=0$, similar to what is done in \cite{azizi,contact terms,khod}, we get the following results corresponding to the emission of the photon from the initial and final mesonic states in terms of form factors:
\begin{eqnarray} \label{first term}
e\varepsilon^{\mu}\varepsilon^{(D_{s1})_{\nu}}
f_{D_{s1}}m_{D_{s1}}T_{\mu\nu}^{(B_{c})}&=&ef_{D_{s1}}m_{D_{s1}}\Bigg\{\Bigg[\Big(\varepsilon.\varepsilon^{(D_{s1})}\Big)(p.q)
-(\varepsilon.p)\Big(\varepsilon^{(D_{s1})}.q\Big)\Bigg]iF_{A}^{(B_{c})}\nonumber
\\&&+if_{B_{c}}\Big(\varepsilon.\varepsilon^{(D_{s1})}\Big)
+\varepsilon_{\nu\mu\lambda\sigma}\varepsilon^{(D_{s1})_{\nu}}\varepsilon^{\mu}p^{\lambda}q^{\sigma}F_{V}^{(B_{c})}\Bigg\},
\end{eqnarray}
\begin{eqnarray} \label{second term final}
ie\varepsilon^{\mu}(p+q)^{\nu}f_{B_{c}}T_{\mu\nu}^{(D_{s1})}&=&ief_{B_{c}}\Bigg\{\Bigg[\Big(\varepsilon.\varepsilon^{(D_{s1})}\Big)(p.q)
-(\varepsilon.p)\Big(\varepsilon^{(D_{s1})}.q\Big)\Bigg]iF_{A}^{(D_{s1})}\nonumber
\\&+&f_{D_{s1}}m_{D_{s1}}\Big(\varepsilon.\varepsilon^{(D_{s1})}\Big)
+\varepsilon_{\nu\mu\lambda\sigma}\varepsilon^{(D_{s1})_{\nu}}\varepsilon^{\mu}p^{\lambda}q^{\sigma}F_{V}^{(D_{s1})}\Bigg\},~~
\end{eqnarray}
where $F_{V(A)}^{(B_{c})}$ and $F_{V(A)}^{(D_{s1})}$ are the transition  form factors. Using Eqs (\ref{first term}),  (\ref{second term final}) and (\ref{2}) we find the WA transition amplitude to be
\begin{eqnarray}\label{3}
&&M^{WA}(B_{c}\rightarrow D_{s1}\gamma) =\notag \\&& e\frac{G_{F}}{\sqrt{2}}V_{cb}V_{cs}^{*}\Bigg(-f_{D_{s1}}m_{D_{s1}}\Bigg\{\Bigg[\Big(\varepsilon.\varepsilon^{(D_{s1})}\Big)(p.q)-(\varepsilon.p)\Big(\varepsilon^{(D_{s1})}.q\Big)\Bigg]iF_{A}^{(B_{c})}
+if_{B_{c}}\Big(\varepsilon.\varepsilon^{(D_{s1})}\Big) \notag \\ &&\ +\varepsilon_{\nu\mu\lambda\sigma}\varepsilon^{(D_{s1})\nu}\varepsilon^{\mu}p^{\lambda}q^{\sigma}F_{V}^{(B_{c})}\Bigg\}
 -if_{B_{c}}\Bigg\{\Bigg[\Big(\varepsilon.\varepsilon^{(D_{s1})}\Big)(p.q)-(\varepsilon.p)\Big(\varepsilon^{(D_{s1})}.q\Big)\Bigg]iF_{A}^{(D_{s1})}\notag \\ &&\ +f_{D_{s1}}m_{D_{s1}}
\Big(\varepsilon.\varepsilon^{(D_{s1})}\Big)
+\varepsilon_{\nu\mu\lambda\sigma}\varepsilon^{(D_{s1})\nu}\varepsilon^{\mu}p^{\lambda}q^{\sigma}F_{V}^{(D_{s1})}\Bigg\}\Bigg).
\end{eqnarray}
As  mentioned in Section I,
the form factors $F_{V}^{(B_{c})}$ and $F_{A}^{(B_{c})}$ 
are calculated in \cite{Aliev1}, so what remains to be calculated  are the form factors $F_{V}^{(D_{s1})}$ and $F_{A}^{(D_{s1})}$, which we discuss  in the  next Section.

\section{QCD SUM RULES For the form factors $F_{V}^{(D_{s1})}$ and $F_{A}^{(D_{s1})}$ \label{Bound}}
To calculate the transition form factors   $F_{V}^{(D_{s1})}$ and $F_{A}^{(D_{s1})}$ via QCD sum rules formalism, we start considering the following correlation function:
\begin{equation}
\Pi_{\mu\nu}(p,q)=i\int d^{4}xe^{iQx}\langle \gamma(q)|T\Big\{\overline{c}(x)\gamma_{\mu}(1-\gamma_{5})s(x)\overline{s}(0)\gamma_{\nu}\gamma_{5}c(0)\Big\}|0\rangle,  \label{CF1}
\end{equation}
where $Q=p+q$. The basic idea in this method is to calculate this correlation function first in hadronic language, called phenomenological or physical side and second in terms of the QCD degrees of freedom using
the operator product expansion in deep Euclidean space, called the theoretical or QCD side. The two representations are then matched in order to get the QCD sum rules for the form factors. To suppress the contributions
coming from the higher energy states and continuum we apply a Borel transformation as well as continuum subtraction which bring two auxiliary parameters: namely the Borel mass parameter and the continuum
 threshold. We shall find their working regions requiring that the physical observables be independent of these parameters.

 First, we focus on calculation of  the phenomenological side. For this aim,  we insert a full set of hadronic  $D_{s1}$ state into Eq.(\ref{CF1}) and perform the four-integral over $x$ to get
\begin{equation}\label{CF2}
 \Pi_{\mu\nu}(p,q)=\frac{\langle \gamma(q)|\overline{c}\gamma_{\mu}(1-\gamma_{5})s|D_{s1}(p)\rangle\langle D_{s1}(p)|\overline{s}\gamma_{\nu}\gamma_{5}c|0\rangle}{m_{D_{s1}}^{2}-p^2}.
\end{equation}
The matrix element, $\langle D_{s1}(p)|\overline{s}\gamma_{\nu}\gamma_{5}c|0\rangle$  is defined in terms of the decay constant and the polarization vector of the $D_{s1}$ meson as
\begin{equation}\label{Mtrx1}
\langle D_{s1}(p)|\overline{s}\gamma_{\nu}\gamma_{5}c|0\rangle=f_{D_{s1}}m_{D_{s1}}\varepsilon_{\nu}^{(D_{s1})},
\end{equation}
while the transition matrix element is parametrized in terms of form factors,
\begin{eqnarray}\label{Mtrx2}
    \langle \gamma(q)|\overline{c}\gamma_{\mu}(1-\gamma_{5})s|D_{s1}(p)\rangle &=&
 e\Bigg\{i\varepsilon_{\mu\alpha\beta\sigma}\varepsilon^{\alpha}\varepsilon^{(D_{s1})\beta}q^{\sigma}\frac{F_{V}^{(D_{s1})}(Q^2)}{m_{D_{s1}}^2} \notag \\ &&\
+\Big[\varepsilon_{\mu}(\varepsilon^{(D_{s1})}.q)-(\varepsilon.\varepsilon^{(D_{s1})})q_{\mu}\Big]\frac{F_{A}^{(D_{s1})}(Q^2)}{m_{(D_{s1})}^2}\Bigg\}.
\end{eqnarray}
Substituting Eqs. (\ref{Mtrx1}) and (\ref{Mtrx2}) into Eq. (\ref{CF2})  and  summing over the polarization vector of the $D_{s1}$ meson, we find the following result for the phenomenological part of the correlation function:
\begin{eqnarray}\label{CF4}
   \Pi_{\mu\nu}(p,q) &=& \frac{ef_{D_{s1}}m_{D_{s1}}}{m_{D_{s1}}^2-p^2}\Bigg\{i\varepsilon_{\mu\nu\alpha\sigma}\varepsilon^{\alpha}q^{\sigma}\frac{F_{V}^{(D_{s1})}(Q^2)}{m_{D_{s1}}^2}
+\Big[q_{\mu}\varepsilon_{\nu}-\varepsilon_{\mu}q_{\nu}\Big]\frac{F_{A}^{(D_{s1})}(Q^2)}{m_{D_{s1}}^2}\Bigg\}.
\end{eqnarray}

We now compute the QCD side of the correlation function within the deep Euclidean region in terms of the QCD parameters. We start by writing the correlation function in terms of the two selected structures as
\begin{equation}\label{CF5}
    \Pi_{\mu\nu}(p,q)=i\varepsilon_{\mu\nu\alpha\sigma}\varepsilon^{\alpha}q^{\sigma}\Pi_{1}+\Big[q_{\mu}\varepsilon_{\nu}-\varepsilon_{\mu}q_{\nu}\Big]\Pi_{2},
\end{equation}
where each function $\Pi_{i}~(i=1$ or $2)$ has  perturbative and non-perturbative parts, i.e.,
\begin{equation}
 \Pi_{i}=\Pi_{i}^{pert}+\Pi_{i}^{non-pert}.
\end{equation}
To calculate  the perturbative parts, we consider diagrams (a) and (b) in figure 2 where the photon can be radiated from both the charm and strange quarks. For the non-perturbative parts, we take into account the
quark condensate and mixed diagrams [diagrams 2 (c), 2 (d) and 2 (e)] as well as diagram 2 (f) for the interaction of the photon  with the soft quark.

The perturbative part in each case  can be written via the dispersion relation as
\begin{equation}\label{Coef1}
    \Pi_{i }^{pert}=\int ds\frac{\rho_{i}(s,Q^2)}{s-p^2} + \textrm{subtraction terms}.
\end{equation}
where $\rho_{i}$ are the spectral densities. Our main task  is now to calculate these spectral densities using the diagrams (a) and (b) in figure 2.
Here, we use a method based on both Feynman and Schwinger parameterizations  with several Borel transformations (see also \cite{Nester}). The Feynman amplitude for the diagram (a) can be written as
\begin{equation}\label{CFa}
    \Pi_{\mu\nu,(a)}=eN_{c}Q_{s}\int \frac{d^{4}k}{(2\pi)^4}\Bigg\{Tr\Big[\frac{i(\not\!k+m_{c})}{k^2+m_{c}^2}\gamma_{\nu}\gamma_{5}
\frac{i(\not\!p+\not\!k+m_{s})}{(p+k)^2-m_{s}^2}\not\!\varepsilon\frac{i(\not\!Q+\not\!k+m_{s})}{(Q+k)^2-m_{s}^2}\gamma_{\mu}(1-\gamma_{5})\Big]\Bigg\},
\end{equation}
where $	N_c=3$ is the number of colors and $Q_s$ is the charge of the strange quark.
\begin{figure}[tbp]
\centering {\includegraphics[width=15cm]{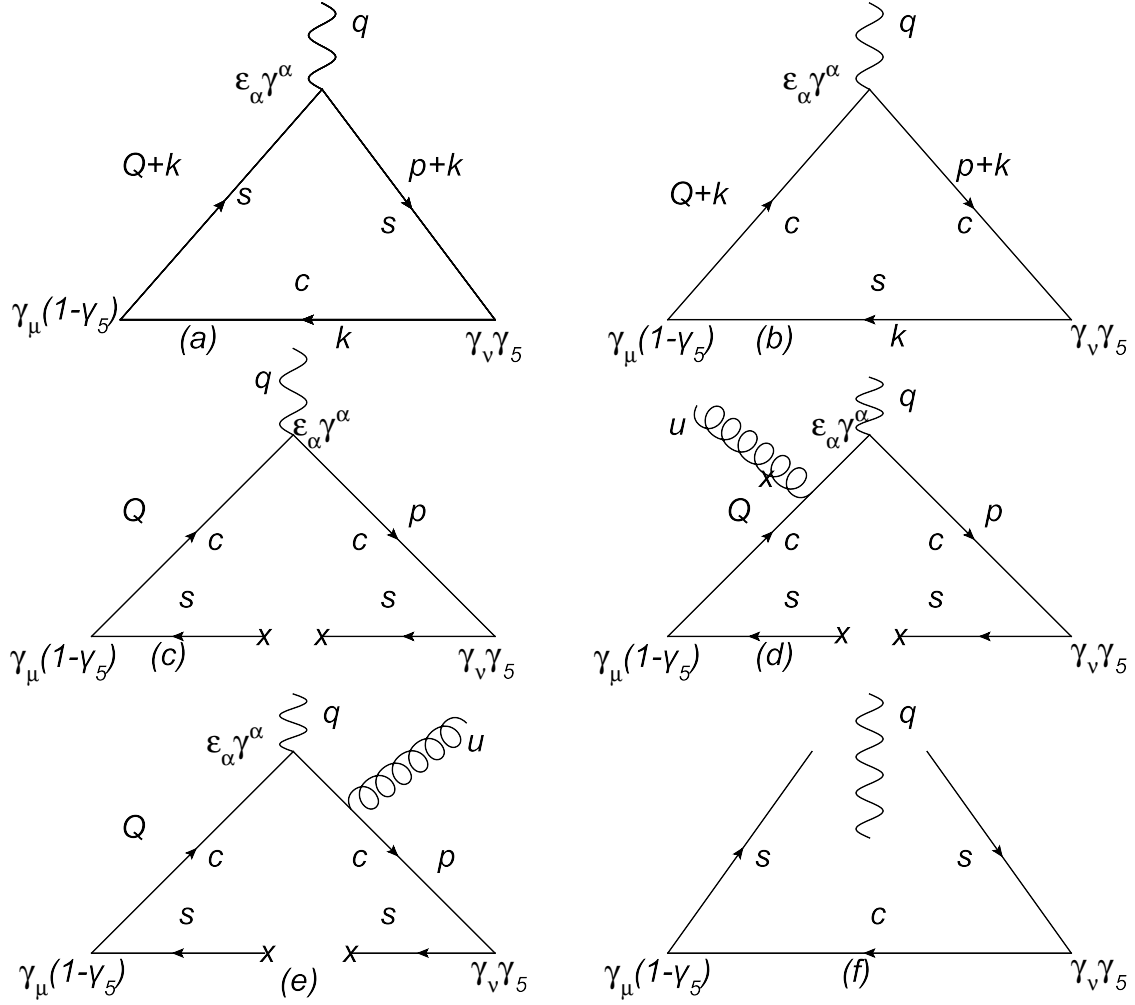}}
\caption{Diagrams for bare-loop [(a), (b)], quark and mixed condensates  [(c), (d), (e)] and propagation of the soft quark in the electromagnetic field (f).}
\label{F2}
\end{figure}
Using  the Feynman parameterization, we perform the four-integral over $k$ and then we use the Schwinger parameterization
\begin{equation}\label{Den}
    \frac{1}{\Delta^n}=\frac{1}{\Gamma(n)}\int_{0}^{\infty}d\alpha \alpha^{n-1} e^{-\alpha \Delta},
  \end{equation}
to write the denominators  in  exponential forms. As a result, we get
\begin{equation}\label{T1}
    \Pi_{1,a }^{pert} = \frac{eN_{c}Q_{s}}{4 \pi^2}\Bigg\{\int_{0}^{1}dxx\int_{0}^{1} dy
   \Big [m_{s} (m_{c} + m_{s} x y) + p^2 x\overline{x} (1-\overline{x} y) +
    2 p.q \overline{x} x^2 y^2 \Big]\int_{0}^{\infty} d\alpha e^{-\alpha \Delta}\Bigg\},
\end{equation}
\begin{equation}\label{T2}
    \Pi_{2 ,a}^{pert} =
 \frac{eNcQ_{s}}{4 \pi^2}\Bigg\{\int_{0}^{1}dxx\int_{0}^{1} dy
   x\Big[m_{s}(m_{c}-m_{s}xy) -
    p^{2} x\overline{x}(1 - \overline{x} y) - 2 p.q \overline{x} x^2 y^2\Big]\int_{0}^{\infty} d\alpha e^{-\alpha \Delta}\Bigg\},
\end{equation}
where $\overline{x}(\overline{y})=1-x(y)$, and $\Delta =
 m_ {c}^2 \overline{x} + m_ {s}^2 x - p^2  x\overline{x} ~\overline{y} -
  Q^2  x\overline{x} y $.

  Applying a double Borel transformation $\widehat{B}(M_{1}^{2})\widehat{B}(M_{2}^{2})$ on $\Pi_{i }^{pert}$, that transforms $Q^{2} \to M_{1}^{2}$ and $p^{2} \to M_{2}^{2}$, we obtain
  \begin{eqnarray}\label{TT1}
    \widehat{\Pi}_{1,a }^{pert} &=& \frac{ eN_ {c} Q_ {s} }{4 \pi^2}
  \frac{ \sigma_ {1} \sigma_ {2}}{\sigma_ {1} + \sigma_ {2}}\int_{0}^{1}dx\frac{1}{\overline{x}}e^{\frac{(m_{c}^2 \overline{x} + m_{s}^2 x)(\sigma_{1} + \sigma_{2})}{\overline{x}x}} \notag \\ &&\ \times\Bigg\{ m_{c} m_{s} +m_{s}^{2} x \frac{\sigma_{1}}{\sigma_{1} + \sigma_{2}} + 2x (1-x^2) \frac{\sigma_{1}}{(\sigma_{1}+\sigma_{2})^2}\Bigg\},\nonumber\\
  \end{eqnarray}
 \begin{eqnarray}\label{TT2}
    \widehat{\Pi}_{2,a }^{pert} &=& \frac{eN_ {c} Q_ {s}}{
  4 \pi^2}\frac{\sigma_{1} \sigma_{2}}{\sigma_{1} + \sigma_{2}}\int_{0}^{1}dx\frac{1}{\overline{x}}
  e^{\frac{(m_ {c}^2 \overline{x} + m_ {s}^2 x) (\sigma_{1} + \sigma_{2})}{
   \overline{x} x}}\notag \\ &&\ \times\Bigg\{m_{c} m_{s} -m_{s}^{2} x \frac{\sigma_{1}}{\sigma_{1} + \sigma_{2}}-2x (1-x^2) \frac{\sigma_{1}}{(\sigma_{1}+\sigma_{2})^2}\Bigg\},\nonumber\\
 \end{eqnarray}
 where $\sigma_{1,2}=1/M_{1,2}^2$ and  we have used  
 \begin{eqnarray}\label{Borel1}
    \widehat{B}_{p^2}(M^2)e^{-\alpha p^2}&=&\delta (1-\alpha M^2),\nonumber\\
 \widehat{B}_{p^2}(M^2)p^2e^{-\alpha p^2}&=&-\frac{d}{d\alpha}\widehat{B}_{p^2}(M^2)e^{-\alpha p^2}=-\frac{d}{d\alpha}\delta (1-\alpha M^2).
 \end{eqnarray}

 Now, we perform a second double Borel transformation on $\widehat{\Pi}_ {i}^{pert}$ in order to transform   $\sigma_{1}$ and $\sigma_{2}$ to the new variables $w$ and $s$ using
 \begin{equation}\label{Spc1}
    \varrho_{i}(w,s)=\frac{1}{ws}\widehat{B}\Big(\frac{1}{w},\sigma_{1}\Big)\widehat{B}\Big(\frac{1}{s},\sigma_{2}\Big)\frac{ \widehat{\Pi}_ {i}^{pert}}{\sigma_{1} \sigma_{2}}.
 \end{equation}
In the calculations, we also use the relations
\begin{eqnarray}\label{Borel3}
   \widehat{B}\Big(\frac{1}{w},\sigma_{1}\Big)\widehat{B}\Big(\frac{1}{s},\sigma_{2}\Big)e^{-\alpha (\sigma_{1}+\sigma_{2})}
&=&\delta (1-\frac{\alpha}{w})\delta(1-\frac{\alpha}{s}),
\end{eqnarray}
and
\begin{eqnarray}
     \sigma^{n}e^{-\alpha \sigma}&=&(-\frac{d}{d\alpha})^{n}e^{- \alpha \sigma}.
 \end{eqnarray}
The final expressions for the spectral densities are then calculated via the following formula:
 \begin{equation}\label{Sp2}
    \rho_{i}(s,Q^2)=\int dw\frac{\varrho_{i}(w,s)}{w-Q^2}.
 \end{equation}

After lengthy calculations, we get the following spectral densities corresponding to the diagram (a):
\begin{eqnarray}\label{Sp31}
    \rho_{1a}(s,Q^2) &=& \frac{eN_{c}Q_{s}}{16\pi^2}\frac{1}{(s-Q^2)^2}\int_{x_{0}}^{x_{1}}dx \frac{1}{x\overline{x}^2}\Bigg\{
  m_c^4 \overline{x}^2(x-5)+m_s^4 x^2 (x-6) -m_c^2 m_s^2  x\overline{x}(2x-11)  \notag \\ &&\  + 4 m_c m_s x \overline{x} (s-Q^2)+
    m_c^2 x\overline{x}^2 \Big[(\overline{x}-8) Q^2 + \overline{x} (s-Q^2)\Big]-
 m_s^2  x^2\overline{x} \Big[(x-10) Q^2 \notag \\ &&\ + (x-2) (s-Q^2)\Big] -4 x^2 \overline{x}^2  Q^2(s-Q^2)  -4  x^2 \overline{x}^2 Q^4\Bigg\},\nonumber\\
\end{eqnarray}
\begin{eqnarray}\label{Sp32}
    \rho_{2a}(s,Q^2) &=& \frac{eN_{c}Q_{s}}{16\pi^2}\frac{1}{(s-Q^2)^2}\int_{x_{0}}^{x_{1}}dx \frac{1}{x\overline{x}^2}\Bigg\{-
  m_c^4 \overline{x}^2(x-5)-m_s^4 x^2(-6 + x)+m_c^2 m_s^2  x\overline{x}(2x-11)\notag \\ &&\ + 4  m_c m_s x \overline{x}(s-Q^2)-
    m_c^2 x \overline{x}^2\Big[ (\overline{x}-8)Q^2 + \overline{x}(s-Q^2)\Big]  +
 m_s^2 x^2\overline{x} \Big[(x-10) Q^2  \notag \\ &&\ + (x-2)(s-Q^2)\Big] +4 x^2 \overline{x}^2  Q^2(s-Q^2) +4  x^2 \overline{x}^2 Q^4   \Bigg\},
\end{eqnarray}
where the integral boundaries $x_{0}$ and $x_{1}$ satisfy  the following inequality:
\begin{equation}\label{ineq}
    sx\overline{x}-(m_{c}^2\overline{x}+m_{s}^2x)\geq0,
\end{equation}
which comes from the definition of the Heaviside-Theta function arising in these calculations.
Similarly, we calculate the contribution of the diagram 2(b). The final expressions for the spectral densities corresponding to the two selected structures are 
\begin{eqnarray}\label{Sp41}
    \rho_{1}(s,Q^2)&=&\frac{eN_{c}}{32\pi^2}\frac{1}{(s-Q^2)^2} \Bigg\{Q_s \Bigg[
  \lambda  \Bigg(4(5 \alpha - 5 \beta-1)sQ^2 +
       s^2 \Big[3  \alpha (\alpha-3)-  \beta(6 \alpha+1) +
          3 \beta^2\Big]\Bigg) \notag \\ &&\ +2 \Bigg(4 m_c m_s (s-Q^2) +8\alpha s Q^2+
    \alpha (1 - 4\alpha + 9\beta)s^2\Bigg) \ln\Big(\frac{
   1 + \alpha - \beta - \lambda}{
   1 + \alpha - \beta + \lambda}\Big)\Bigg] \notag \\ &&\ + Q_c \Bigg[
  \lambda  \Bigg(4(5 \beta - 5 \alpha-1)sQ^2 +
       s^2 \Big[3  \beta (\beta-3)-  \alpha(6 \beta+1) +
          3 \alpha^2\Big]\Bigg) \notag \\ &&\ +2 \Bigg(4 m_c m_s (s-Q^2) +8\beta s Q^2+
    \beta (1 - 4\beta + 9\alpha)s^2\Bigg) \ln\Big(\frac{
   1 + \beta - \alpha - \lambda}{
   1 + \beta - \alpha + \lambda}\Big)\Bigg]\Bigg\},
\end{eqnarray}
\begin{eqnarray}\label{Sp42}
    \rho_{2}(s,Q^2)&=&\frac{eN_{c}}{32\pi^2}\frac{1}{(s-Q^2)^2}\Bigg\{Q_s \Bigg[ \lambda \Bigg(4(- 5 \alpha + 5 \beta+1)s Q^2  +
    s^2 \Big[
       -3\alpha(\alpha-3)+
          \beta(6 \alpha+1)- 3\beta^2\Big]\Bigg) \notag \\ &&\  +
 2 \Bigg(4 m_c m_s (s-Q^2) - 8 s \alpha  Q^2 -
     \alpha (1 - 4  \alpha + 9 \beta)s^2\Bigg) \ln\Big(\frac{
   1 + \alpha - \beta - \lambda}{
   1 + \alpha - \beta + \lambda}\Big)\Bigg] \notag \\ &&\  +Q_c \Bigg[ \lambda \Bigg(4(- 5 \beta + 5 \alpha+1)s Q^2  +
    s^2 \Big[
       -3\beta(\beta-3)+
          \alpha(6 \beta+1)- 3\alpha^2\Big]\Bigg) \notag \\ &&\  +
 2 \Bigg(4 m_c m_s (s-Q^2) - 8 s \beta  Q^2 -
     \beta (1 - 4  \beta + 9 \alpha)s^2\Bigg) \ln\Big(\frac{
   1 + \beta - \alpha - \lambda}{
   1 + \beta - \alpha + \lambda}\Big)\Bigg] \Bigg\},
\end{eqnarray}
where $\alpha=\frac{m_{s}^2}{s}$, $\beta=\frac{m_{c}^2}{s}$ and $\lambda=\sqrt{1+\alpha^2+\beta^2-2\alpha-2\beta-2\alpha \beta}$.

For the non-perturbative parts,   we begin by calculatting  contributions of the quark condensate and mixed diagrams ( diagrams (c), (d) and (e) in figure 2) and  obtain 
\begin{eqnarray}\label{nonp1}
    \Pi_{1(c,d,e)}^{non-pert} &=& \frac{m_{c}}{r^2 R^2}\langle\overline{s}s\rangle + \frac{m_{s}}{2}\langle\overline{s}s\rangle\Big[
  \frac{2}{r^2 R^2} +  \frac{m_{c}^2}{ r^4 R^2} - \frac{7 m_{c}^2}{ r^2 R^4} - \frac{4  m_{c}^4}{
   r^4 R^4}\Big]\notag \\ &&\ + \frac{m_{s}^2}{
  2}\langle\overline{s}s\rangle\Big[\frac{ 2 m_{c}^3}{r^2 R^6} - \frac{8 m_{c}^5}{r^6 R^4} + \frac{2 m_{c}^3}{r^4 R^4} - \frac{
   3 m_{c}}{r^2 R^4} + \frac{2 m_{c}^3}{r^6 R^2} - \frac{m_{c}}{r^4 R^2}\Big]\notag \\ &&\ + \frac{
  m_{0}^2}{
  12}\langle\overline{s}s\rangle\Big[\frac{- 6 m_{c}^3}{r^2 R^6} + \frac{24  m_{c}^5}{r^6 R^4} - \frac{
   6 m_{c}^3}{r^4 R^4} + \frac{8 m_{c}}{r^2 R^4} - \frac{6 m_{c}^3}{
   r^6 R^2} + \frac{3 m_{c}}{r^4 R^2}\Big] - \frac{4  m_{c}^3}{r^2 R^4}\langle\overline{s}s\rangle,\nonumber\\
\end{eqnarray}
\begin{eqnarray}\label{nonp2}
    \Pi_{2(c,d,e)}^{non-pert} &=&-\frac{m_{c}}{r^2 R^2}\langle\overline{s}s\rangle + \frac{m_{s}}{
  2}\langle\overline{s}s\rangle\Big[\frac{1}{r^2 R^2} + \frac{1}{R^4} - \frac{4 m_{c}^4}{r^4 R^4} - \frac{3 m_{c}^2}{
   r^2 R^4} + \frac{m_{c}^2}{2 r^4 R^2}\Big]\notag \\ &&\ + \frac{m_{s}^2}{
  2}\langle\overline{s}s\rangle\Big[-\frac{2 m_{c}^3}{r^2 R^6} + \frac{8 m_{c}^5}{r^6 R^4} - \frac{2 m_{c}^3}{
   r^4 R^4} + \frac{3 m_{c} }{ r^2 R^4} - \frac{2 m_{c}^3}{r^6 R^2} + \frac{m_{c}}{
   r^4 R^2}\Big]\notag \\ &&\ + \frac{m_{0}^2}{
  4}\langle\overline{s}s\rangle\Big[\frac{2 m_{c}^3}{ r^2 R^6} - \frac{8 m_{c}^5}{r^6 R^4} + \frac{2 m_{c}^3}{
   r^4 R^4} - \frac{4 m_{c}}{r^2 R^4} + \frac{2 m_{c}^3}{r^6 R^2} -  \frac{m_{c} }{
   r^4 R^2}\Big] + \frac{4 m_{c}^3}{r^2 R^4}\langle\overline{s}s\rangle,\nonumber\\
\end{eqnarray}
where $r^2=p^2-m_{c}^2$ and $R^2=Q^2-m_{c}^2$.

The final contribution to the WA mode is that of diagram (f). This diagram corresponds to the propagation of the soft quark in the external electromagnetic field.
Here we need to make use of the  light-cone version of the QCD sum rules and photon distribution amplitudes (DAs). The relevant correlation function is of the form:
\begin{equation}\label{CFLC1bir}
    \Pi_{\mu\nu,(f)}(p,q)=i\int d^{4}x e^{-iQx} \langle\gamma(q)|T\Big\{\overline{s}(0)\gamma_{\mu}\gamma_{5}c(0)\overline{c}(x)\gamma_{\nu}(1-\gamma_{5})s(x)\Big\}|0\rangle.
\end{equation}
Contracting the $c$-quark lines in Eq.(\ref{CFLC1bir}) and using the propagator of the heavy quark in momentum space, we obtain
\begin{equation}\label{CFLC2}
    \Pi_{\mu\nu,(f)}(p,q)=i^2\int d^{4}x \frac{d^{4}k}{(2\pi)^4}\frac{e^{-i(Q-k)x}}{m_{c}^2-k^2} \langle\gamma(q)|\overline{s}\gamma_{\mu}\gamma_{5}(\not\!k+m_{c})\gamma_{\nu}(1-\gamma_{5})s|0\rangle.
\end{equation}
To relate the matrix element  in the above equation to the photon DAs, we use the identities
\begin{eqnarray}\label{Iden}
    \gamma_{\mu}\gamma_{\nu} &=& g_{\mu\nu}+i\sigma_{\mu\nu}, \notag \\
    \gamma_{\mu}\gamma_{\nu}\gamma_{5} &=& g_{\mu\nu}\gamma_{5}-\frac{i}{2}\varepsilon_{\mu\nu\alpha\beta}\sigma_{\alpha\beta}\notag, \\
    \gamma_{\mu}\gamma_{\alpha}\gamma_{\nu} &=& g_{\mu\alpha}\gamma_{\nu}+g_{\nu\alpha}\gamma_{\mu}-g_{\mu\nu}\gamma_{\alpha}+i\varepsilon_{\mu\nu\alpha\lambda}\gamma_{\lambda}\gamma_{5}.
\end{eqnarray}
The relevant photon DAs of twist 2, 3, and 4 \cite{DA1, DA2} are
\begin{eqnarray}\label{DA}
    \langle\gamma(q)|\overline{s}\gamma_{\nu}s|0\rangle &=& -\frac{Q_{s}}{2}f_{3\gamma}\int_{0}^{1}du \overline{\psi}^{(V)}(u)x^{\theta}F_{\theta\nu}(u x), \notag \\
    \langle\gamma(q)|\overline{s}\gamma_{\alpha}\gamma_{5}s|0\rangle &=& -\frac{iQ_{s}}{4}f_{3\gamma}\int_{0}^{1}du \overline{\psi}^{(A)}(u)x^{\theta}\widetilde{F}_{\theta\alpha}(u x), \notag \\
    \langle\gamma(q)|\overline{s}\sigma_{\alpha\beta}s|0\rangle &=& Q_{s}\langle\overline{s}s\rangle\int_{0}^{1}du\phi(u)F_{\alpha\beta}(ux)\notag \\ &&\
    +\frac{Q_{s}\langle\overline{s}s\rangle}{16}\int_{0}^{1}du x^{2}\textbf{A}(u)F_{\alpha\beta}(ux)\notag \\ &&\
    +\frac{Q_{s}\langle\overline{s}s\rangle}{8}\int_{0}^{1}du \textbf{B}(u)x^{\rho}(x_{\beta}F_{\alpha\rho}(ux)-x_{\alpha}F_{\beta\rho}),
\end{eqnarray}
where $F_{\mu\nu}$ is the field strength tensor of the electromagnetic field and is defined by
\begin{eqnarray}\label{Fmunu}
    F_{\mu\nu}(x) &=& -i(\varepsilon_{\mu}q_{\nu}-\varepsilon_{\nu}q_{\mu})e^{iqx},
\end{eqnarray}
and
\begin{eqnarray}
    \widetilde{F}_{\mu\nu}(x) &=& \frac{1}{2}\varepsilon_{\mu\nu\alpha\beta}F_{\alpha\beta}(x).
\end{eqnarray}
The wave function $\phi(u)$ is defined in terms of the magnetic susceptibility $\chi(\mu)$ at a renormalization scale ($\mu=1\textrm{GeV}^{2}$) in the following manner:
\begin{equation}\label{WavFun1}
    \phi(u)=\chi(\mu)u(1-u).
\end{equation}
The remaining functions $\overline{\psi}^{(V)}(u)$, $\overline{\psi}^{(A)}(u)$, $\textbf{A}(u)$, and $\textbf{B}(u)$ are also defined as \cite{DA1, DA2}
\begin{eqnarray}\label{WavFun2}
    \overline{\psi}^{(V)}(u) &=& -20u(1-u)(2u-1)+\frac{15}{16}(\omega_{\gamma}^{A}-3\omega_{\gamma}^{V})u(1-u)(2u-1)(7(2u-1)^2-3),\notag \\
    \overline{\psi}^{(A)}(u) &=& (1-(2u-1)^2)(5(2u-1)^2-1)\frac{5}{2}(1+\frac{19}{16}\omega_{\gamma}^{V}-\frac{3}{16}\omega_{\gamma}^{A}),\notag \\
    \textbf{A}(u) &=& 40u(1-u)(3k-k^{+}+1)+8(\xi_{2}^{+}-3\xi_{2})\notag \\ &&\
    \times[u(1-u)(2+13u(1-u))+2u^3(10-15u+16u^2)\ln u\notag \\ &&\
    +2(1-u)^3(10-15(1-u)+6(1-u^2))\ln(1-u)],\notag \\
    \textbf{B}(u) &=& 40\int_{0}^{u}d\alpha(4-\alpha)(1+3k^{+})[-\frac{1}{2}+\frac{3}{2}(2\alpha-1)^2],
\end{eqnarray}
where $k$, $k^{+}$, $\xi_{2}$, $\xi_{2}^{+}$ and $f_{3\gamma}$ are constants (see \cite{DA1, DA2}).
Putting the above equations all together and after performing the four-integrals over $x$ and $k$, the coefficients of the corresponding structures,
 $i\varepsilon_{\mu\nu\alpha\beta}\varepsilon^{\alpha}q^{\beta}$ and $[q_{\mu}\varepsilon_{\nu}-\varepsilon_{\mu}q_{\nu}]$ are obtained as follows:
\begin{eqnarray}\label{LCST}
    \Pi_{1f}^{non-pert}(p,q) &=& \frac{Q_{s}}{
 2 (m_{c}^2 -
    p^2)^3}\int_{0}^{1}du\Big\{m_{c}^3 \langle\overline{s}s\rangle\textbf{A}(u)  \notag \\ &&\
     + (m_{c}^2 - p^2) \Big[m_{c} \langle\overline{s}s\rangle\textbf{B}(u)  -
      2 (5 m_{c}^2 - p^2) (m_{c} \langle\overline{s}s\rangle \phi(u) - f_{3\gamma} \psi^{(V)}(u))\Big]\Big\},\notag \\
\end{eqnarray}
\begin{eqnarray}
      \Pi_{2f}^{non-pert}(p,q) &=& \frac{m_{c} Q_{s}}{
 2 (m_{c}^2 -
    p^2)^3}\int_{0}^{1}du \Big\{\textbf{A}(u) m_{c}^2  \langle\overline{s}s\rangle\ +
  2 (-5 m_{c}^2 + p^2)  \langle\overline{s}s\rangle\ \phi(u)\notag \\ &&\
   + (m_{c}^2 - p^2) \Big[\textbf{B}(u)  \langle\overline{s}s\rangle\ +
     f_{3\gamma} m_{c} \psi^{(A)}(u)\Big]\Big\}.
\end{eqnarray}

Now, to find the QCD sum rules for the form factors  we match the coefficients of the selected structures from both phenomenological and QCD sides and perform the
Borel transformation with respect to the momentum of $D_{s1}$ meson $(p^{2} \to M_{B}^{2})$. To further suppress the contributions of the higher energy states and continuum we also perform the continuum subtraction
 and use the quark-hadron duality assumption and find
\begin{equation}\label{FFWeak}
    F_{V,A}^{(D_{s1})}(Q^2)=\frac{m_{D_{s1}}}{f_{D_{s1}}} e^{m_{D_{s1}}/M_{B}^2}\widehat{B}\Bigg\{\int_{(m_{s}+m_{c})^2}^{s_{0}}ds\frac{\rho_{1,2}(s,Q^2)}{s-p^2}+\Pi_{1,2(c+d+e+f)}^{non-pert}\Bigg\},
\end{equation}
where $s_0$ is the continuum threshold and the $V$  ($A$) on the left-hand side corresponds to the 1 (2) on the right-hand side.
To obtain the expressions for the above sum rules in the Borel scheme, we perform the Borel transformation using the  standard rule
\begin{equation}\label{B2}
    \widehat{B}\frac{1}{(p^{2}-s)^{n}}=(-1)^{n}\frac{e^{-s/M_{B}^{2}}}{\Gamma(n)(M_{B}^{2})^{n-1}}.
\end{equation}

\section{ QCD SUM RULES FOR THE FORM FACTORS RESPONSIBLE FOR THE  ELECTROMAGNETIC PENGUIN MODE\label{Bound}}
At the quark level, the FCNC based EP transition of the $B_{c} \to D_{s1}\gamma$  proceeds via  $b \to s \gamma$ whose effective Hamiltonian is written as
\begin{equation}\label{Hamil}
    H^{eff}=-\frac{G_{F}e}{4\pi^2\sqrt{2}}V_{tb}V_{ts}^{*}C_{7}(\mu)\overline{s}\sigma_{\mu\nu}\Big[m_{b}\frac{1+\gamma_{5}}{2}+m_{s}\frac{1-\gamma_{5}}{2}\Big]bF^{\mu\nu}.
\end{equation}
The amplitude of this mode is obtained from
\begin{equation}
 M^{EP}=\langle D_{s1}(p)|H^{eff}|B_{c}(Q)\rangle,
\end{equation}
hence to proceed further, we need to calculate the following matrix elements:
\begin{equation}\label{MatrixCL}
    \langle D_{s1}|\overline{s}\sigma_{\mu\nu}(1\pm\gamma_{5})q^{\nu}b|B_{c}\rangle,
\end{equation}
which can be parametrized in terms of two gauge invariant form factors $T_{1}(q^{2})$ and $T_{2}(q^{2})$ in the case of real photon, i.e.
\begin{eqnarray}\label{Matrix}
    \langle D_{s1}(p,\varepsilon^{(D_{s1})})|\overline{s}\sigma_{\mu\nu}q^{\nu}\gamma_{5}b|B_{c}(Q)\rangle &=& i\varepsilon_{\mu\alpha\beta\lambda}\varepsilon^{(D_{s1})\alpha}p^{\beta}Q^{\lambda}T_{1}(0),\notag \\
    \langle D_{s1}(p,\varepsilon^{(D_{s1})})|\overline{s}\sigma_{\mu\nu}q^{\nu}b|B_{c}(Q)\rangle &=&
    \Big[(m_{B_{c}}^2-m_{D_{s1}}^2)\varepsilon_{\mu}^{(D_{s1})}-(\varepsilon^{(D_{s1})}.q)(p+Q)_{\mu}\Big]T_{2}(0),\nonumber\\
\end{eqnarray}
where these two from factors are not independent from each other. Using the relation, $\sigma_{\mu\nu}\gamma_{5}=-\frac{i}{2}\varepsilon_{\mu\nu\alpha\beta}\sigma^{\alpha\beta}$, we see
$T_{1}(0)=\frac{1}{2}T_{2}(0)$. Therefor, we need to calculated just one of them, and  we choose to calculate the form factor $T_{2}(0)$. The corresponding  correlation function is chosen as
\begin{equation}\label{CFP1}
    \Pi_{\mu\nu}(p^{2},Q^{2})=i^2\int\int d^{4}x d^{4}y e^{-i(Qx-py)}\langle0|T\Big\{\overline{c}(y)\gamma_{\nu}\gamma_{5}s(y)\overline{s}(0)\sigma_{\mu\alpha}q^{\alpha}b(0)\overline{b}(x)\gamma_{5}c(x)\Big\}|0\rangle,
\end{equation}
where $\overline{b}\gamma_{5}c$ and  $\overline{c}\gamma_{\nu}\gamma_{5}s$ are the interpolating currents of the initial and final mesonic states, respectively, and $\overline{s}\sigma_{\mu\alpha}q^{\alpha}b$
is the transition current.
Using the general philosophy of the QCD sum rules we calculate this correlation function  in two different languages: namely the hadronic language and the quark-gluon language. For the hadronic, or phenomenological side, we get
\begin{eqnarray}\label{CFLC1}
    \Pi_{\mu\nu} (p^{2},Q^{2})&=& \frac{if_{D_{s1}}f_{B_{c}}m_{D_{s1}}m_{B_{c}}^{2}}{(m_{B_{c}}^{2}-Q^{2})(m_{D_{s1}}^{2}-p^{2})(m_{b}+m_{c})}\Bigg\{(m_{B_{c}}^{2}-m_{D_{s1}}^{2})g_{\mu\nu}T_{2}(0) \notag \\ &&\
    -\Big(\frac{m_{B_{c}^{2}}-m_{D_{s1}}^{2}}{m_{D_{s1}}^{2}}\Big)p_{\mu}p_{\nu}T_{2}(0)+(p+Q)_{\mu}\Big[\frac{p.q}{m_{D_{s1}}^{2}}p_{\nu}-q_{\nu}\Big]T_{2}(0)\Bigg\}+....,
\end{eqnarray}
where $...$ denotes contributions of the higher energy states and continuum which will be suppressed by applying the Borel transformation as well as the continuum subtraction,
 and we have used the
following definition of the decay constant of the $B_c$ meson:
\begin{equation}
 \langle B_c|\overline{b}\gamma_{5}c|0\rangle=i\frac{f_{B_c}m_{B_c}^2}{(m_b+m_c)}.
\end{equation}
To calculate the form factor $T_2(0)$,   we choose the structure $g_{\mu\nu}$.

In the QCD side, the correlation function is  written in terms of the selected structure as
\begin{equation}\label{CFLC2}
    \Pi_{\mu\nu}=g_{\mu\nu}\Pi(p^{2},Q^{2}),
\end{equation}
where
\begin{equation}
 \Pi(p^{2},Q^{2})=\Pi^{pert}(p^{2},Q^{2})+\Pi^{non-pert}(p^{2},Q^{2}).
\end{equation}
Here, the perturbative part is related to the spectral density, $\rho^{pert}(s',t)$ by a double dispersion integral,
\begin{equation}\label{StPer}
    \Pi^{pert}(p^2,Q^2)=-\frac{1}{(2\pi)^2}\int \int ds' dt \frac{\rho^{pert}(s',t)}{(s'-Q^2)(t-p^2)}+\mathrm{\textrm{subtraction terms}},
\end{equation}
and for the non-perturbative contributions we will calculate the two-gluon condensate diagrams.

 Now, we focus our attention on calculating the spectral density.
Using the Cutkosky method \cite{cutkbey}, we get
\begin{eqnarray}\label{PertPen}
    \rho^{\mathrm{per}}(s',t) &=& 2 Nc\Big\{I_{0} \Big[\Delta [(m_{b} - m_{c}) (m_{c} + m_{s}) + t] -
      \Delta' [(m_{b} - m_{c}) (m_{c} + m_{s}) + s'] \notag \\ &&\  +
      2 m_{c} [(m_{c} + m_{s}) s' + m_{b} t - m_{c} t] - m_{c} (m_{b} + m_{s}) u\Big] +
   2 A1 (-2 s' + u)\Big\},
\end{eqnarray}
where
\begin{eqnarray}\label{I0AB}
    I_{0} &=& \frac{1}{4 \sqrt{\lambda'(s', t, q^2)}},\notag \\
\lambda'(a,b,c)&=&a^2+b^2+c^2-2ab-2ac-2bc,\notag \\
    A_{1} &=& \frac{-I_{0}}{(-4 s' t + u^2)^2} [\Delta^2 t+\Delta'^2 s'-\Delta \Delta'u+m_{c}^2 (-4 s' t + u^2)],\notag \\
    \Delta &=& s' + m_{c}^2 - m_{b}^2,\notag \\
    \Delta' &=& t + m_{c}^2 - m_{s}^2,\notag \\
    u &=& t + s' - q^2.
\end{eqnarray}
Note that, to obtain the above spectral density, we have performed the integrals over the delta-functions which restricts the boundaries of the integrals on the  $s'$ and $t$ as:
\begin{eqnarray}\label{ineq}
    m_{c}^2\leq t\leq t_{0},\notag \\
    t-\frac{t m_{b}^2}{m_{c}^2-t}\leq s'\leq s'_{0},
\end{eqnarray}
where $s'_0$ and $t_0$ are the continuum thresholds in the initial and final channels in the case of EP  mode.
\begin{figure}[tbp]
\centering {\includegraphics[width=15cm]{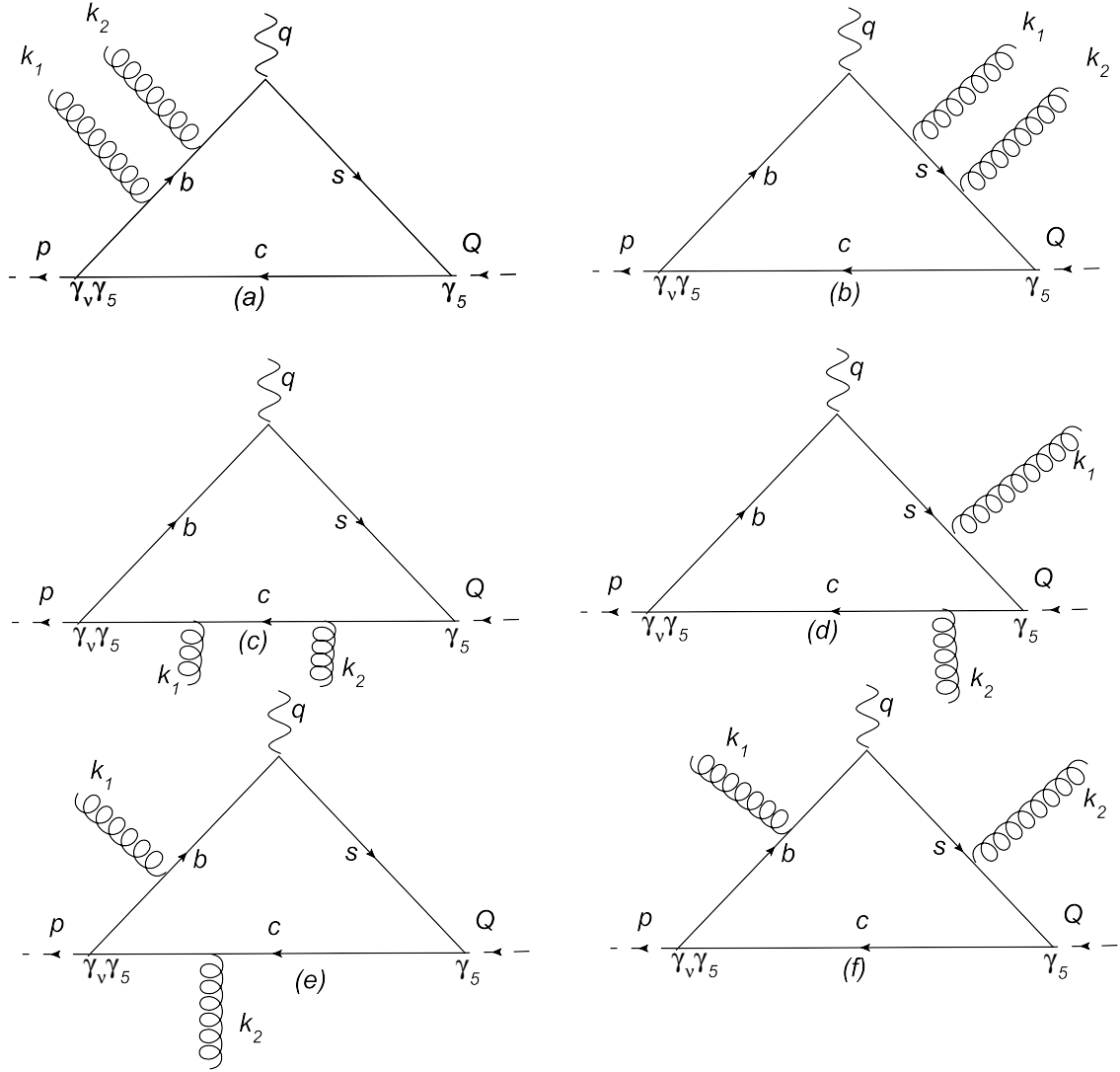}}
\caption{Feynman diagrams for gluon condensates corrections.}
\label{Fgluon}
\end{figure}
There are several sources for non-perturbative contributions, such as quark, quark-gluon, and gluon condensates, however, the quark-quark and quark-gluon condensates give
zero contributions after applying the double Borel transformation with respect to $Q^{2}$ ($Q^{2} \to M_{1}^{2}$) and $p^{2}$ ($p^{2} \to M_{2}^{2}$). Therefore, the
remaining source of the non-perturbative contributions would be the gluon condensates [see Fig(3)]. The calculation of such contributions is lengthy but standard. For the non-perturbative part in the Borel scheme, we get
\begin{equation}
 \Pi^{non-pert}=M_{1}^{2}M_{2}^{2}\Big\langle\frac{\alpha_{s}}{\pi}G^{2}\Big\rangle C_{G^{2}},
\end{equation}
where $C_{G^{2}}$ is the Wilson coefficient of the gluon condensates that is defined as
\begin{equation}\label{CG2}
    C_{G^{2}}=C_{G^{2}}^{a}+C_{G^{2}}^{b}+C_{G^{2}}^{c}+C_{G^{2}}^{d}+C_{G^{2}}^{e}+C_{G^{2}}^{f}.
\end{equation}
The explicit expressions of the $C_{G^{2}}^{i}$ are given in the Appendix.

Using a similar procedure  to that presented in the previous section, we get the  sum rule  for the form factor $ T_{2}(0)$ to be 
\begin{eqnarray}\label{FormFactLC}
    T_{2}(0) &=& \frac{(m_{b}+m_{c})e^{m_{B_{c}}^{2}/M_{1}^{2}}e^{m_{D_{s1}}^{2}/M_{2}^{2}}}{if_{D_{s1}}f_{B_{c}}m_{D_{s1}}m_{B_{c}}^{2}(m_{B_{c}}^{2}-m_{D_{s1}}^{2})}
\Bigg[\frac{-1}{(2\pi)^{2}}\int \int ds' dt e^{-s'/M_{1}^{2}}e^{-t/M_{2}^{2}}\rho^{\textrm{pert}}(s',t) \notag \\ &&\
    +M_{1}^{2}M_{2}^{2}\Big\langle\frac{\alpha_{s}}{\pi}G^{2}\Big\rangle C_{G^{2}}\Bigg].
\end{eqnarray}

\section{NUMERICAL ANALYSIS\label{Bound}}
This section is devoted to the numerical analysis of the form factors,  estimating the branching ratio in the  WA and EP channels and  the total branching fraction of the $B_{c}\rightarrow D_{s1}(2460)\gamma$ transition . For this aim, we use the
quark and mesons' masses as  $m_{c}=(1.275\pm 0.015)~GeV$, $m_{s}\simeq142~MeV$  \cite{Ref23}, $m_{b}=(4.7\pm 0.1)~GeV$  \cite{Ref22}, $m_{D_{s1}}=(2459.6\pm0.6)~MeV$, $m_{B_{c}}=(6.277\pm0.006)~GeV$  \cite{PDG}.
 For the values of the  decay constants, we use  $f_{D_{s1}}=(225 \pm 25)~MeV$ and
$f_{B_{c}}=(350\pm25)~MeV$  \cite{Ref24,Ref25,Ref26}. The values of the condensates are \cite{Ref22}: $\langle\overline{\psi}\psi|_{\mu=1GeV}\rangle=-(240 \pm 10 MeV)^{3}$,
$\langle\overline{s}s\rangle=(0.8 \pm 0.2)\langle\overline{\psi}\psi\rangle$, $m_{0}^{2}=(0.8\pm0.2)~GeV^{2}$ and
  $\langle\frac{\alpha_{s}}{\pi}G^{2}\rangle=(0.012\pm0.004)~GeV^4$. The  parameters entered the photon DAs are also taken as
  $\chi=(3.15 \pm 0.30)~ GeV^{-2}$, $k=0.2$, $k^{+}=0$, $\zeta_{1}=0.4$, $\zeta_{1}^{+}=0$, $\zeta_{2}=0.3$, $\zeta_{2}^{+}=0$, $f_{3\gamma}=-(4 \pm 2)\times 10^{-3} GeV^{2}$,
 $\omega_{\gamma}^{A}=-2.1 \pm 1.0$ and $\omega_{\gamma}^{V}=3.8 \pm 1.8$ \cite{DA1, DA2, Ref28}. The remaining parameters are chosen as
 $|V_{cs}|=0.957 \pm 0.017$, $|V_{cb}|=0.0416 \pm 0.0006$,
 $|V_{tb}|=0.77_{-0.24}^{+0.18}$, $|V_{ts}|=(40.6 \pm 2.7) \times 10^{-3}$ \cite{PDG}, $C_{7}(\mu=m_{c})=-0.0068-0.02i$ \cite{Greub},
 and $\tau_{B_{c}}=0.52 \times 10^{-12} s$.

The sum rules for the form factors contain also the  continuum thresholds and the Borel mass parameters as auxiliary objects. We  find  working regions  for these parameters such that the physical observables are practically independent
of them. The continuum thresholds are not completely arbitrary but  are correlated with the energy of the first excited states in the initial and final mesonic channels.
Our numerical results show that the results depend weakly on the thresholds in the intervals
 $s_{0}= t_0=(6-8) GeV^{2}$ and $s'_{0}=(45-50) GeV^{2}$. The working regions for the Borel parameters are obtained by demanding that not only the contributions of the higher states and continuum are effectively
suppressed, but also the contributions of the higher order operators and higher twist DAs are small, i.e., the series of the sum rules  converge. These conditions lead to the intervals,
 $6~ GeV^{2}\leq M_{B}^{2} \leq 12~ GeV^2$, $10~ GeV^{2}\leq M_{1}^{2} \leq 30~ GeV^{2}$
and $5~ GeV^{2}\leq M_{2}^{2} \leq 12 ~GeV^{2}$ for the Borel mass parameters.

Now, we proceed to find the fit functions of the form factors using the aforesaid working regions for the auxiliary parameters.
Here we would like to mention that, for the decay rates, we need only the values of the form factors $F_{V}^{(D_{s1})}$ and $F_{A}^{(D_{s1})}$ at $Q^2=m_{B_{c}}^2$,  $F_{V}^{(B_{c})}$ and $F_{A}^{(B_{c})}$
 at $p^2=m_{D_{s1}}^2$ and  $T_2$ at $q^2=0$. However, we determine their fit functions in general and   give  their values at these fixed points.  The fit functions for the  form factors
$F_{V}^{(D_{s1})}$ and $F_{A}^{(D_{s1})}$
 are 
\begin{equation}\label{FitFun}
    f(Q^2)=\frac{f(0)}{1+a\frac{Q^2}{m_{D_{s1}}^{2}}+b(\frac{Q^2}{m_{D_{s1}}^{2}})^2},
\end{equation}
where $f(0)$, $a$ and $b$ are the fit parameters whose values are 

\begin{eqnarray*}
  \begin{tabular}{|c|c|c|c|}\hline
$\textrm{form factors}$ &  $f(0)$     &$a$   & $b$  \\ \hline
$F_{V}^{(D_{s1})}(Q^2)$   &  $0.098$   &$0.171$ & $-0.008 $  \\
$F_{A}^{(D_{s1})}(Q^2)$   &  $-2.478$   &$3.644$ & $-0.005$ \\
\hline
\end{tabular}
\end{eqnarray*}
The values of these form factors at $Q^2=m_{B_{c}}^2$ are 
\begin{eqnarray}\label{FDs1}
    F_{V}^{(D_{s1})}(Q^2=m_{B_{c}}^2) &=& 0.055 \pm0.016 ,\notag \\
    F_{A}^{(D_{s1})}(Q^2=m_{B_{c}}^2) &=&-0.102\pm0.030,
\end{eqnarray}
where the errors on the values are due to the uncertainties in determination of the working regions for the auxiliary parameters as well as those coming from the DAs and other input parameters.

Also the fit functions for the form factors $F_{A,V}^{(B_{c})}$ are \cite{Aliev1}
\begin{eqnarray}\label{FitBc}
    F_{V}^{(B_{c})}(p^2) &=& \frac{F_{V}(0)}{1-p^2/m_{1}^2}, \notag \\
    F_{A}^{(B_{c})}(p^2) &=& \frac{F_{A}(0)}{1-p^2/m_{2}^2},
\end{eqnarray}
where the fit parameters are 
\begin{eqnarray*}
  \begin{tabular}{|c|c|}\hline

$F_{V}(0)=0.44 \textrm{GeV}$   &  $m_{1}^2=43.10 \textrm{GeV}^2$ \\\hline
$F_{A}(0)=0.21 \textrm{GeV}$   &  $m_{2}^2=48.00 \textrm{GeV}^2$ \\\hline
\end{tabular}
\end{eqnarray*}
The values of the form factors $F_{V,A}^{(B_{c})}$ calculated at $p^2=m_{D_{s1}}^2$ are 
\begin{eqnarray}\label{FBc}
    F_{V}^{(B_{c})}(p^2=m_{D_{s1}}^2) &=& (0.51\pm0.14) \textrm{GeV} ,\notag \\
    F_{A}^{(B_{c})}(p^2=m_{D_{s1}}^2) &=& (0.24 \pm0.07) \textrm{GeV}.
\end{eqnarray}

For the form factor induced by the EP at $q^2=0$, we obtain
\begin{equation}\label{T2}
    T_{2}(0) = -0.298\pm0.085.
\end{equation}

At the end of this section we would like to calculate the  decay widths and branching ratios. Using the amplitudes of each decay mode, we find the following expressions for the  decay rates
at fixed points  in WA and EP channels as well as for the total decay rate of the transition under consideration:
 \begin{eqnarray}\label{BrWA}
    \Gamma^{(WA)}(B_{c}\to D_{s1} \gamma) &=&  \frac{G_{F}^{2} \alpha |V_{cb}V^*_{cs}|^{2}}{16}\Big(\frac{m_{B_{c}}^{2}-m_{D_{s1}}^{2}}{m_{B_{c}}}\Big)^{3}
 \notag \\ &&\ \times\Bigg\{f_{B_{c}}^2 \Big[(F_{A}^{(D_{s1})})^{2} + (F_{V}^{(D_{s1})})^{2}\Big]+2 f_{B_{c}} f_{D_{s1}} F_{V}^{(B_{c})} F_{V}^{(D_{s1})}\frac{m_{D_{s1}}}{m_{B_{c}}^{2}}
\notag \\ &&\ +f_{D_{s1}}^{2} m_{D_{s1}}^{2}\Big[\frac{(F_{A}^{(B_{c})})^{2}}{m_{B_{c}}^{4}} +\frac{(F_{V}^{(B_{c})})^{2}}{m_{B_{c}}^{4}}\Big]\Bigg\},
 \end{eqnarray}
 \begin{eqnarray}\label{BrEP}
      \Gamma^{(EP)}(B_{c}\to D_{s1} \gamma) &=& \frac{G_{F}^{2} \alpha |C_{7}|^{2} |V_{tb}V^*_{ts}|^{2}}{1024 \pi^{4}}\Big(\frac{m_{B_{c}}^{2}-m_{D_{s1}}^{2}}{m_{B_{c}}}\Big)^{3}
 \notag \\ &&\ \times \Big(16(m_b+m_s)^2+(m_b-m_s)^{2}\Big)[T_{2}(0)]^{2},
 \end{eqnarray}
 \begin{eqnarray}\label{Brtot}
   \Gamma^{(total)}(B_{c}\to D_{s1} \gamma) &=& \frac{G_{F}^2 \alpha}{
 1024\pi^4} \Big(\frac{m_{B_{c}}^2 - m_{D_{s1}}^2}{
  m_{B_{c}}}\Big)^3 \Bigg\{64 \pi^4 |V_{cb}V^*_{cs}|^2 \Bigg[f_{B_{c}}^2 \Big\{(F_{A}^{(D_{s1})})^2 + (F_{V}^{(D_{s1})})^2\Big\} \notag \\ &&\ +
      2 f_{B_{c}} f_{D_{s1}} F_{V}^{(B_{c})} F_{V}^{(D_{s1})}  \frac{m_{D_{s1}}}{m_{B_{c}}^2} +
      f_{D_{s1}}^2 m_{D_{s1}}^2 \Big\{\frac{(F_{A}^{(B_{c})})^2}{m_{B_{c}}^4} + \frac{(F_{V}^{(B_{c})})^2}{m_{B_{c}}^4}\Big\}\Bigg] \notag \\ &&\ +
   |C_{7}|^2 |V_{tb}V^*_{ts}|^2 \Big(16(m_b-m_s)^2+(m_b+m_s)^{2}\Big) [T_{2}(0)]^2  \notag \\ &&\ + 16\pi^2 T_{2}(0) |V_{cb}V^*_{cs}| |V_{tb}V^*_{ts}| \Bigg[ f_{D_{s1}} m_{D_{s1}}
 X \Big\{4\frac{F_{A}^{(B_{c})}}{m_{B_{c}}^2} (m_{b} - m_{s})  \notag \\ &&\ + \frac{F_{V}^{(Bc)}}{m_{B_{c}}^2} (m_{b} + m_{s})\Big\}  +
       f_{B_{c}} \Big\{F_{V}^{(D_{s1})} (m_{b} + m_{s}) X  \notag \\ &&\ -4 F_{A}^{(D_{s1})} (m_{b} - m_{s}) Y\Big\}\Bigg]\Bigg\},
 \end{eqnarray}
where $X$ and $Y$ are the real and imaginary parts of the Wilson coefficient $C_{7}$, respectively. In these formulas,  the fixed point values  of the form factors  are used.


Finally the numerical values of the corresponding branching ratios for the radiative decay under consideration are obtained as follows:
\begin{eqnarray}\label{Br}
    \textbf{B}^{(EP)}(B_{c}\to D_{s1} \gamma) &=& (1.769 \pm 0.582)\times 10^{-8}, \notag \\
    \textbf{B}^{(WA)}(B_{c}\to D_{s1} \gamma)&=& (2.243 \pm 0.736) \times 10^{-5}, \notag \\
    \textbf{B}^{(total)}(B_{c}\to D_{s1} \gamma) &=& (2.351 \pm 0.795) \times 10^{-5},
\end{eqnarray}
where the dominant contribution to each channel comes from the perturbative part. From these values, we also see that the $B_{c}\rightarrow D_{s1}(2460)\gamma$ transition
proceeds mostly via the WA mode.
The order of the total branching ratio indicates that this decay channel can be detected at LHCb in near future. Any measurement on this decay and the comparison of the obtained data with our predictions in the present work
can give valuable information about the nature and internal structure of the participating particles, especially the $D_{s1}$ meson.


\acknowledgements

One of the authors (A. R. O.) would like to thank R. Khosravi and S. Zarepour for useful discussions. Also partial support of Shiraz university research council is appreciated.

\section{APPENDIX\label{Bound}}

The explicit expressions for $C_{G^2}^{i}$ are given as follows:
\begin{eqnarray}\label{CGa}
    C_{G^2}^{a} &=& m_{b}\Bigg\{-2m_{b}m_{c}^2 [I_{0}[1,3,1]-3m_{b}^2I_{0}[1,4,1]+2m_{b}^2 (m_{b}-m_{s})(m_{b}+m_{s})I_{0}[1,5,1]] \notag \\ &&\ +2m_{c}(I_{0}[1,2,1]-4m_{b}^2 I_{0}[1,3,1]-m_{b}m_{s}I_{0}[1,3,1]+5m_{b}^4 I_{0}[1,4,1]\notag \\ &&\ +3m_{b}^3 m_{s}I_{0}[1,4,1]-3m_{b}^2 m_{s}^2 I_{0}[1,4,1]+2m_{b}^3 (-m_{b}+m_{s})(m_{b} +m_{s})^2 I_{0}[1,5,1])\notag \\ &&\ +2m_{s}[I_{0}[1,2,1]-2m_{b}^6 I_{0}[1,5,1]+m_{b}^4 (5I_{0}[1,4,1]+2I_{0}^{[0,1]}[1,5,1])  \notag \\ &&\ +m_{b}^2 (-4I_{0}[1,3,1]-3I_{0}^{[0,1]}[1,4,1]+3I_{0}^{[1,0]}[1,3,1])] \notag \\ &&\ +m_{b}(3I_{0}^{[0,1]}[1,3,1]+I_{0}^{[0,2]}[1,4,1]+I_{0}^{[1,0]}[1,3,1]+2m_{b}^4 (3I_{0}^{[0,1]}[1,5,1]+I_{0}^{[1,0]}[1,5,1]) \notag \\ &&\ -m_{b}^2 (9I_{0}^{[0,1]}[1,4,1]+2I_{0}^{[0,2]}[1,5,1]+3I_{0}^{[1,0]}[1,4,1]-2I_{0}^{[2,0]}[1,4,1]) \notag \\ &&\ -4I_{3}^{[0,1]}[1,5,1]+4I_{3}^{[1,0]}[1,3,1])\Bigg\},
\end{eqnarray}
\begin{eqnarray}\label{CGb}
    C_{G^2}^{b} &=& 7m_{b}^2 m_{c}^2 m_{s}^2 I_{0}[1,1,3]+m_{b}^3 I_{0}[1,1,4]+m_{c}I_{0}[1,1,4]+m_{b}m_{c}I_{0}[1,1,4]  \notag \\ &&\ -m_{b}^2 m_{c}I_{0}[1,1,4]-m_{b}^3 m_{s}I_{0}[1,1,4]-m_{c}m_{s}I_{0}[1,1,4]+2m_{b}m_{c}m_{s}I_{0}[1,1,4]  \notag \\ &&\ +m_{b}^2 m_{c}m_{s}I_{0}[1,1,4] -m_{c}I_{0}[1,1,5]-m_{b}m_{c}I_{0}[1,1,5]+m_{b}^3 m_{c}I_{0}[1,1,5]-m_{b}^2 m_{c}^2 I_{0}[1,1,5]  \notag \\ &&\ +m_{c}m_{s}I_{0}[1,1,5]-2m_{b}m_{c}m_{s}I_{0}[1,1,5]+2m_{b}^3 m_{c}m_{s}I_{0}[1,1,5]-2m_{b}^2 m_{c}^2 m_{s}I_{0}[1,1,5]  \notag \\ &&\ -m_{b}I_{0}^{[0,1]}[1,1,4]+m_{s}I_{0}^{[0,1]}[1,1,4]-3/2 I_{0}^{[0,1]}[1,1,5]+3/2m_{b}^2 I_{0}^{[0,1]}[1,1,5]  \notag \\ &&\ +m_{b}m_{s}I_{0}^{[0,1]}[1,1,5]-m_{s}I_{0}^{[0,2]}[1,1,2]+3/2 I_{0}^{[1,0]}[1,1,3] +m_{b}I_{0}^{[1,0]}[1,1,3] \notag \\ &&\ +3m_{s}I_{0}^{[1,0]}[1,1,3]+1/2 m_{b}^2 I_{0}^{[1,0]}[1,1,4]-m_{b}m_{s}I_{0}^{[1,0]}[1,1,4]+1/2 I_{0}^{[2,0]}[1,1,4] \notag \\ &&\ +m_{s}I_{0}^{[2,0]}[1,1,4],
\end{eqnarray}
\begin{eqnarray}\label{CGc}
    C_{G^2}^{c} &=& 1/6\Bigg\{2m_{b}^5 [m_{s}(-I_{0}[3,1,1]+m_{c}^2 I_{0}[3,1,2]+I_{0}^{[0,1]}[3,1,1]) \notag \\ &&\ +m_{c}(m_{c}^2 I_{0}[3,2,2]+I_{0}^{[0,1]}[3,1,2])]-3I_{0}^{[0,1]}[3,2,1]+I_{0}^{[0,1]}[3,2,2]+3I_{0}^{[0,2]}[3,1,2]  \notag \\ &&\ -2I_{0}^{[0,2]}[3,2,1]
    +I_{0}^{[0,3]}[3,2,2]+3I_{0}^{[1,0]}[3,1,1]-I_{0}^{[1,0]}[3,1,2]  \notag \\ &&\ +2m_{c}^3 m_{s}\{I_{0}[3,2,1]-I_{0}[3,2,2]-4I_{0}^{[0,1]}[3,1,2]+3I_{0}^{[0,1]}[3,2,1]+3I_{0}^{[1,0]}[3,1,1] \notag \\ &&\-2I_{0}^{[1,0]}[3,2,1]\}
    -I_{0}^{[1,1]}[3,1,1]+I_{0}^{[1,1]}[3,1,2]-2m_{b}^3 [-m_{c}(-2I_{0}^{[0,1]}[3,2,2]+I_{0}^{[1,0]}[3,1,2] \notag \\ &&\ +m_{c}^2 (I_{0}[3,1,2]-2I_{0}[3,2,2]-3(I_{0}^{[0,1]}[3,2,2]+I_{0}^{[1,0]}[3,2,2]))+I_{0}^{[1,1]}[3,2,1])  \notag \\ &&\
    +m_{s}((1+2m_{c}^2)I_{0}[3,2,2]+2I_{0}^{[0,1]}[3,1,1]-2I_{0}^{[0,1]}[3,2,1]+4m_{c}^2 I_{0}^{[0,1]}[3,2,2] \notag \\ &&\ +I_{0}^{[0,2]}[3,1,2]+I_{0}^{[2,1]}[3,1,2]))-m_{b}^2 (2m_{c}^6 I_{0}[3,2,2]+I_{0}^{[0,1]}[3,1,2]-6I_{0}^{[0,1]}[3,2,1] \notag \\ &&\ -2I_{0}^{[0,2]}[3,1,2] +6I_{0}^{[0,2]}[3,2,2]+I_{0}^{[0,3]}[3,2,2]+2m_{c}^3 m_{s}(I_{0}[3,1,2]-5I_{0}^{[0,1]}[3,2,2] \notag \\ &&\ -I_{0}^{[1,0]}[3,2,1])+3I_{0}^{[1,0]}[3,2,1]-2I_{0}^{[1,0]}[3,2,2]  -7m_{c}^4 I_{0}^{[1,0]}[3,2,2] \notag \\ &&\ +2m_{c}m_{s}(I_{0}[3,1,2]-I_{0}[3,2,1]+2I_{0}^{[0,2]}[3,1,2]-I_{0}^{[1,1]}[3,2,1]) +2I_{0}^{[1,1]}[3,2,1] \notag \\ &&\ -3I_{0}^{[1,2]}[3,2,2]+2I_{0}^{[2,0]}[3,2,1]-2I_{0}^{[2,1]}[3,2,2])+I_{0}^{[2,1]}[3,2,2]  \notag \\ &&\ +m_{c}^2 (-2I_{0}^{[0,1]}[3,2,2] +10I_{0}^{[0,2]}[3,1,2]-2I_{0}^{[0,2]}[3,2,1]+I_{0}^{[1,0]}[3,2,1]+I_{0}^{[1,0]}[3,2,2] \notag \\ &&\ -14I_{0}^{[1,1]}[3,1,2]+14I_{0}^{[1,1]}[3,2,2]+3I_{0}^{[1,2]}[3,2,2]+2I_{0}^{[2,0]}[3,1,2] -10I_{0}^{[2,0]}[3,2,1] \notag \\ &&\ -3I_{0}^{[2,1]}[3,2,2]-3I_{0}^{[3,0]}[3,2,2])-I_{0}^{[3,0]}[3,2,2]\Bigg\},
\end{eqnarray}
\begin{eqnarray}\label{CGd}
    C_{G^2}^{d} &=& 1/12\Bigg\{2m_{c}^4I_{0}[3,2,1]+2m_{c}^3 m_{s}I_{0}[3,1,1]+24m_{b}^7 (m_{c}+m_{s})I_{0}[3,2,2]+I_{0}^{[0,1]}[3,2,2] \notag \\ &&\ +2m_{b}^6 (8m_{c}m_{s}I_{0}[3,1,1]+8m_{c}^2 I_{0}[3,2,2]  +3I_{0}^{[0,1]}[3,1,2]+I_{0}^{[1,0]}[3,1,1]) \notag \\ &&\ +6m_{b}^5 (2m_{c}^3 I_{0}[3,1,1]+2m_{c}^2 m_{s}I_{0}[3,2,2]-2m_{s}(-2I_{0}[3,1,2] +2m_{s}^2I_{0}[3,2,1] \notag \\ &&\ +2I_{0}^{[0,1]}[3,2,2])+m_{c}-9I_{0}^{[0,1]}[3,2,1]  +I_{0}^{[1,0]}[3,1,2]))+3I_{0}^{[1,0]}[3,2,1]-I_{0}^{[1,0]}[3,2,2] \notag \\ &&\ -2m_{c}m_{s}(2I_{0}^{[0,1]}[3,1,2]-I_{0}^{[1,0]}[3,1,1]
    -I_{0}^{[1,0]}[3,1,2])+2I_{0}^{[1,1]}[3,2,2]+I_{0}^{[1,2]}[3,1,2] \notag \\ &&\ -m_{c}^2 (-2I_{0}[3,1,2]+2I_{0}[3,2,2]+2I_{0}^{[0,1]}[3,2,2]+I_{0}^{[0,2]}[3,2,2]-5I_{0}^{[1,0]}[3,2,1] \notag \\ &&\ +3I_{0}^{[1,0]}[3,2,2]-I_{0}^{[2,0]}[3,2,2])-I_{0}^{[2,0]}[3,2,2]+m_{b}^4 (4m_{c}^4 I_{0}[3,2,1]+4m_{c}^3 m_{s}I_{0}[3,2,2] \notag \\ &&\ +2I_{0}^{[0,1]}[3,2,1]-15I_{0}^{[0,1]}[3,2,2]-2I_{0}^{[0,2]}[3,1,1]    +5I_{0}^{[0,1]}[3,2,2]-4I_{0}^{[1,0]}[3,1,2]) \notag \\ &&\ +6I_{0}^{[1,0]}[3,2,2] +3(I_{0}^{[0,1]}[3,2,2]+I_{0}^{[1,0]}[3,2,2])+6I_{0}^{[1,1]}[3,1,2]+4I_{0}^{[2,0]}[3,2,2])
    \notag \\ &&\  +3m_{b}(2m_{c}^2 m_{s}(-I_{0}[3,2,1]-2I_{0}^{[0,1]}[3,2,2]+2I_{0}^{[1,0]}[3,2,2])-4I_{0}[3,2,2] \notag \\ &&\  +3I_{0}^{[0,1]}[3,2,1]
     -9I_{0}^{[0,1]}[3,2,2]-3I_{0}^{[0,2]}[3,1,2]+5I_{0}^{[1,0]}[3,1,2]+I_{0}^{[1,0]}[3,2,1] \notag \\ &&\  +3I_{0}^{[2,0]}[3,1,2])
     -4m_{s}(I_{0}[3,1,2]+I_{0}^{[0,1]}[3,2,2]+I_{0}^{[1,0]}[3,1,2]-2I_{0}^{[1,0]}[3,2,1] \notag \\ &&\ -I_{0}^{[1,1]}[3,2,1]+I_{0}^{[2,0]}[3,2,2]) -3m_{b}^3 (-4m_{c}^3I_{0}[3,2,2]+2m_{c}^2 m_{s}-4I_{0}^{[0,1]}[3,2,2] \notag \\ &&\ +4I_{0}^{[1,0]}[3,1,2]) -4m_{s}(-3I_{0}[3,2,1]+4I_{0}[3,2,2]+3I_{0}^{[0,1]}[3,2,2]-4I_{0}^{[1,0]}[3,2,2] \notag \\ &&\ -2I_{0}^{[1,1]}[3,2,2]
     +2I_{0}^{[2,0]}[3,2,2])+m_{c}(-16I_{0}[3,1,2] +12I_{0}[3,2,2]-27I_{0}^{[0,1]}[3,1,2]\notag \\ &&\ +6I_{0}^{[0,1]}[3,2,1]-6I_{0}^{[0,2]}[3,2,1]+3I_{0}^{[1,0]}[3,1,1] +10I_{0}^{[1,0]}[3,2,2]+6I_{0}^{[2,0]}[3,2,2]) \notag \\ &&\  +m_{b}^2 (m_{c}^4 (-6I_{0}[3,1,2]+4I_{0}[3,2,2])+m_{c}^3 (4m_{s}I_{0}[3,1,2]-6m_{s}I_{0}[3,2,1]) \notag \\ &&\ -3I_{0}^{[0,1]}[3,1,1]+12I_{0}^{[0,1]}[3,1,2]-9I_{0}^{[1,0]}[3,1,2]+4I_{0}^{[1,0]}[3,2,1]
     \notag \\ &&\ +2m_{c}m_{s}(7I_{0}[3,1,1]+9I_{0}^{[0,1]}[3,2,2]-2I_{0}^{[1,0]}[3,2,1]-6I_{0}^{[1,0]}[3,2,2])-9I_{0}^{[1,1]}[3,1,1]
     \notag \\ &&\  -2I_{0}^{[1,2]}[3,2,1]+m_{c}^2 (14I_{0}[3,1,2]-12I_{0}[3,2,2]+9I_{0}^{[0,1]}[3,1,2] \notag \\ &&\ -2I_{0}^{[0,1]}[3,2,2]+2I_{0}^{[0,2]}[3,1,2] -10I_{0}^{[1,0]}[3,1,2]+9I_{0}^{[1,0]}[3,2,1]-2I_{0}^{[2,0]}[3,1,2]) \notag \\ &&\ -6I_{0}^{[2,0]}[3,1,2]+6I_{0}^{[2,0]}[3,2,2]
     +2I_{0}^{[3,0]}[3,1,0])-I_{0}^{[3,0]}[3,2,2]\Bigg\},
\end{eqnarray}
\begin{eqnarray}\label{CGe}
    C_{G^2}^{e} &=& 1/6\Bigg\{-2m_{c}m_{s}(-I_{0}[1,3,2]+I_{0}[1,3,3])+4m_{b}^5 (-m_{c}I_{0}[1,2,3]-2m_{s}I_{0}[1,3,3]) \notag \\ &&\ -2m_{c}^2 [I_{0}[1,2,2]-I_{0}[1,3,3]]-3I_{0}^{[0,1]}[1,1,3]+I_{0}^{[0,1]}[1,2,2]-I_{0}^{[0,2]}[1,2,3] \notag \\ &&\ -I_{0}^{[1,0]}[1,2,2]
    +3I_{0}^{[1,0]}[1,3,1]-2m_{b}^4 (-2m_{c}m_{s}I_{0}[1,2,1]-2m_{c}^2I_{0}[1,3,3]) \notag \\ &&\ +3I_{0}^{[0,1]}[1,3,3]+I_{0}^{[1,0]}[1,2,2])-2m_{b}^3 [m_{c}(-3I_{0}[1,2,3]+2I_{0}[1,3,3]) \notag \\ &&\ +2m_{s}(-I_{0}[1,3,3]-2I_{0}^{[0,1]}[1,2,2]  +2I_{0}^{[1,0]}[1,3,2])] \notag \\ &&\ +2m_{b}\Big\{m_{c}[-I_{0}[1,2,2]+I_{0}[1,3,2]]+2m_{s}[-I_{0}[1,2,3]+I_{0}[1,3,2] \notag \\
  && -I_{0}^{[0,1]}[1,1,3]
    +I_{0}^{[1,0]}[1,2,3]
   -3I_{0}[1,3,1]+2I_{0}^{[0,1]}[1,2,3]-2I_{0}^{[1,0]}[1,3,3])]\Big\}+I_{0}^{[2,0]}[1,3,2] \notag \\ &&\  +m_{b}^2 (2m_{c}^2 (2I_{0}[1,2,3]-3I_{0}[1,3,2])+2m_{c}m_{s}(2I_{0}[1,1,2]-3I_{0}[1,3,3] )+9I_{0}^{[0,1]}[1,2,3] \notag \\ &&\ -2I_{0}^{[0,1]}[1,3,2]+2I_{0}^{[0,2]}[1,3,3]-6I_{0}^{[1,0]}[1,1,3]+3I_{0}^{[1,0]}[1,3,2]
   -2I_{0}^{[2,0]}[1,3,3])\Bigg\},
\end{eqnarray}
\begin{eqnarray}\label{CGf}
    C_{G^2}^{f} &=& 2/3 m_{s}\Bigg\{m_{b}^3 (-m_{c}^2I_{0}[2,1,4])+m_{b}^2 m_{c}(m_{c}^2I_{0}[2,1,4]-2I_{0}^{[0,1]}[2,1,4]) \notag \\ &&\ +m_{b}(I_{0}^{[0,1]}[2,1,3]-2I_{0}^{[0,1]}[2,1,4]
    +I_{0}^{[0,2]}[2,1,2]+I_{0}^{[1,0]}[2,1,3]+m_{c}^2 I_{0}^{[1,0]}[2,1,3] \notag \\ &&\ -I_{0}^{[1,1]}[2,1,4]) +m_{c}(2I_{0}^{[0,1]}[2,1,3]+I_{0}^{[0,2]}[2,1,3]-3I_{0}^{[1,0]}[2,1,4])\Bigg\},
\end{eqnarray}
where we have ignored terms with  higher powers of the strange quark mass. The functions, $I_{n}[a,b,c]$ and $I_{n}^{[i,j]}[a,b,c]$ are defined as:
\begin{eqnarray}\label{I0ij}
    I_{0}[a,b,c] &=& \frac{(-1)^{a+b+c}}{16\pi^2 \Gamma(a) \Gamma(b) \Gamma(c)}(M_{1}^2)^{2-a-b}(M_{2}^2)^{2-a-c}\emph{\textbf{U}}_{0}(a+b+c-4,1-c-b), \notag \\
    I_{1}[a,b,c] &=& \frac{(-1)^{a+b+c+1}}{16\pi^2 \Gamma(a) \Gamma(b) \Gamma(c)}(M_{1}^2)^{2-a-b}(M_{2}^2)^{3-a-c}\emph{\textbf{U}}_{0}(a+b+c-5,1-c-b), \notag \\
    I_{2}[a,b,c] &=& \frac{(-1)^{a+b+c+1}}{16\pi^2 \Gamma(a) \Gamma(b) \Gamma(c)}(M_{1}^2)^{3-a-b}(M_{2}^2)^{2-a-c}\emph{\textbf{U}}_{0}(a+b+c-5,1-c-b), \notag \\
    I_{n}^{[i,j]}[a,b,c] &=& [M_{1}^2]^{i}[M_{2}^2]^{j}\frac{d^{i}}{d(M_{1}^2)^i}\frac{d^{j}}{d(M_{2}^2)^j}[M_{1}^2]^{i}[M_{2}^2]^{j}I_{n}[a,b,c].
\end{eqnarray}
where $\emph{\textbf{U}}_{0}(a,b)$  is given by
\begin{equation}\label{U0}
    \emph{\textbf{U}}_{0}(a,b) = \int_{0}^{1} dy (y+M_{1}^2+M_{2}^2)^{a} y^b \exp[-\frac{B_{-1}}{y}-B_{0}-B_{1}y],
\end{equation}
and
\begin{eqnarray}\label{B}
    B_{-1} &=& \frac{m_{b}^2}{M_{1}^2}[M_{1}^2+M_{2}^2], \notag \\
    B_{0} &=& \frac{1}{M_{1}^2M_{2}^2}[M_{1}^2m_{c}^2+M_{2}^2(m_{c}^2+m_{b}^2)], \notag \\
    B_{1} &=& \frac{m_{c}^2}{M_{1}^2M_{2}^2}.
\end{eqnarray}

\end{document}